# Strain-Gradient–Driven Decoupling of Thermal Suppression from Anisotropy in β-Ga$_2$O$_3$

Guangwu Zhang [a], Xing Xiang [a], Ziyan Qian [a], Yixin Xu [a], Shengying Yue [b], Hyejin Jang [c,d], Lin Yang [e], Yanguang Zhou [a], Xinyu Wang [f,*], Qiye Zheng [a,*]

[a] Department of Mechanical and Aerospace Engineering, The Hong Kong University of Science and Technology, Hong Kong SAR, China

[b] Laboratory for Multiscale Mechanics and Medical Science, SV LAB, School of Aerospace, Xi'an Jiaotong University, Xi'an 710049, China

[c] Department of Materials Science and Engineering, Seoul National University, Seoul 08826, Republic of Korea

[d] Research Institute of Advanced Materials (RIAM), Seoul National University, Seoul 08826, Republic of Korea

[e] School of Advanced Manufacturing and Robotics, Peking University, Beijing 100871, China

[f] Institute of Thermal Science and Technology, Shandong University, Jinan 250061, China

## Abstract

β-Ga$_2$O$_3$ is an emerging ultra-wide-bandgap semiconductor (UWBGS) for efficient, high-frequency power electronics, solar-blind/UV photodetectors, and wearable/flexible devices, with scalable manufacturing. However, its intrinsically low thermal conductivity ($k$)—compounded by ubiquitous, nonuniform strains introduced during fabrication and operation—creates a stringent thermal-management bottleneck that degrades heat dissipation, reliability, and performance. Consequently, understanding how uniform and non-uniform strains affect thermal transport in this UWBG material is essential for thermal management design. Yet, strain gradients ($\eta$), pervasive in flexible devices and epitaxial nanostructures, remain a major blind spot in $\beta$-Ga$_2$O$_3$ thermal transport studies. By integrating the first-principles-based machine learning approach with Boltzmann transport equation, we establish that $\eta$ unlocks a $k$ suppression mechanism fundamentally more potent than uniform strain ($\varepsilon$): moderate uniaxial gradients (0.6%/nm) suppress $k$ by 32–37% (27–30%) in thin films (nanowires), intensifying to 43.3% with biaxial gradients. This reduction far exceeds that from equivalent $\varepsilon$ and surpasses benchmark materials like silicon and BAs. Critically, a surprising decoupling emerges: while $\varepsilon$ ($\pm$3%) alters thermal anisotropy by ~25%, $\eta$ strongly suppresses $k$ with preserving this ratio. Mechanistically, $\eta$-induced symmetry breaking and enhanced mode coupling anisotropically activate forbidden scattering channels, making gradient-driven scattering dominant over intrinsic phonon scattering below 6.25 THz. The strong anisotropic thermodynamics and low-lying optical phonons profoundly shape heat-carrying phonons, rendering $k$ highly sensitive at the device level to $\eta$. These findings redefine non-uniform strain from a parasitic flaw into a powerful design tool for engineering thermal isolation and heat flux in next-generation flexible and high-power $\beta$-Ga$_2$O$_3$ electronics.

---

[*] Corresponding authors: *E-mail addresses*: xyw@sdu.edu.cn (X. Wang), qiyezheng@ust.hk (Q. Zheng).

## 1. Introduction

Ultra-wide bandgap semiconductors (UWBGS) are at the forefront of next-generation power electronics and optoelectronic devices, enabling extreme power densities at high frequencies with improved efficiency and reliability [1]. Among these materials, $\beta$-Ga$_2$O$_3$ stands out for its wide bandgap (4.8 eV) and high breakdown electric field (~8 MV/cm) [2], yielding a superior Baliga figure of merit (3214.1), surpassing GaN and SiC [3,4]. Mechanically, $\beta$-Ga$_2$O$_3$ features a moderate Young's modulus ($\approx$201 GPa [5,6], lower than that of GaN and SiC) and remarkable room-temperature plasticity in submicron films [7], giving it intrinsic flexibility. The exceptional electronic and mechanical properties render $\beta$-Ga$_2$O$_3$ a highly promising material for high-power electronics [8–11], solar-blind and UV photodetectors [12,13], and emerging flexible nanodevices [3,14–16].

Despite these advantages, the device deployment of β-Ga$_2$O$_3$ is hindered by its intrinsically low thermal conductivity ($k$, 10–27 W m$^{-1}$ K$^{-1}$, greatly lower than SiC and GaN [17]), creating critical thermal management challenges that compromise device reliability and limit high-frequency/power operation [17–19]. This bottleneck is further compounded by ubiquitous strains ($\varepsilon$) arising from fabrication and operation that significantly alter its thermal transport properties. For instance, heteroepitaxial β-Ga$_2$O$_3$ films grown on relatively high-$k$ substrates like Al$_2$O$_3$ [20], SiC [21], SrTiO$_3$ [22], ZnO [23], MgO [24], GaN [25], and polycrystalline diamond [26] can experience strains up to 4.8% (GaN) and 6.6% (sapphire) [27–31] due to lattice and thermal expansion mismatches. These epitaxial strains are typically biaxial and nonuniform across the film thickness (Fig. 1a), creating significant strain gradients ($\eta$). Similarly, flexible electronic architectures such as freestanding thin-film and nanowire β-Ga$_2$O$_3$ device (Fig. 1b and Fig. 1c) can often experience bending strains from substrate deformation, producing considerable $\eta$ [3,31].

Previous studies of strain effects on thermal transport in Ga$_2$O$_3$ polymorphs have been limited to theoretical predictions under uniform $\varepsilon$ in bulk materials. For instance, in β-Ga$_2$O$_3$, a 2% tensile (-4% compressive) uniaxial strain along the $b$ axis ($\varepsilon^b$) was predicted to decrease (increase) the $b$-axis $k$ ($k_b$) by -19% (50%) [32] changing anisotropy ratio (AR) of $k$ ($k_b/k_a$, $k_b/k_c$, and $k_c/k_a$), while biaxial strains in α- and κ-phases [33] showed even more dramatic effects. However, real devices develop inhomogeneous strains; thin films host nanoscale $\eta$ [34–36] that strongly modify thermal AR [37,38] and heat removal [3,39], causing inaccuracies in device thermal-resistance modeling and thermal-management design [40–42]. Indeed, recent studies in other nanomaterials have shown that $\eta$ can suppress $k$ far more dramatically than $\varepsilon$. For example, experimentally, a moderate uniaxial $\eta$= 0.112%/nm in 35-nm×85-nm Si nanoribbons produced a 34% reduction in $k$—over three times the reduction under equivalent uniform $\varepsilon$—primarily due to phonon spectrum broadening [43]. Molecular dynamics simulations further show that a uniaxial $\eta$=0.44%/nm in silicon nanocubes enhance AR to 1.20 while suppressing $k$ [44]. And in BAs nanoribbons, similar $\eta$ (0.99-4.97%/nm) nearly doubles the $k$ reduction compared to Si [45]. These results argue that $\eta$ must be treated explicitly in crystalline semiconductors.

Unlike these cubic crystals with isotropic $k$, monoclinic (C2/m) β-$Ga_2O_3$ exhibits two nonequivalent Ga sites and three distinct O sites, producing diverse bond lengths/angles [32], and consequently, pronounced anisotropy in phonon anharmonicity and $k$ tensor [46]. Its $k$ is further shaped by low-lying optical phonons, whose interactions with acoustic modes enhance sensitivity to structural perturbations [47]. Therefore, $k$ and AR in β-$Ga_2O_3$ are expected to be especially susceptible to $\eta$ [32]. However, capturing the $\eta$ effect on thermal transport requires substantial computational resources, necessitating a computational framework that achieves first-principles accuracy while maintaining phonon mode resolvability. Although molecular dynamics is a convenient and widely used computational approach, its accuracy is ultimately constrained by the underlying interatomic potential [48,49]. In contrast, machine-learning interatomic potentials constructed from first-principles data have emerged as a powerful route to achieve both accuracy and efficiency. When combined with the Boltzmann transport equation [50,51] (BTE), these first-principles–based machine learning potentials enable reliable and scalable calculations of phonon mode-resolved thermal transport, bridging the gap between atomistic fidelity and tractable computational cost [52].

In this work, we comprehensively investigate thermal transport in β-$Ga_2O_3$ under uniaxial and biaxial $\varepsilon$ and $\eta$ for 10 nm and 100 nm films—dimensions relevant to device applications [3,13,36] and prior experiments [32,33,53]—using first-principles-based neuroevolution potentials [52] (NEP, details in SI note I) combined with phonon BTE. Our results demonstrate $k$ suppression along multiple axes under $\eta$ exceeds that from corresponding uniform tensile $\varepsilon$ in both uniaxial and biaxial cases at 300 K. Detailed analysis of phonon spectral broadening, mode-resolved scattering rates, and phase space reveals that $\eta$ activate previously forbidden scattering channels and generate localized vibrational modes, dramatically enhancing phonon scattering. Notably, $\eta$ has a much weaker effect on thermal anisotropy, in contrast to uniform $\varepsilon$. These results demonstrate that nonuniform strain critically shapes phonon transport in β-$Ga_2O_3$ and underscore $\eta$ as an essential consideration for accurate thermal modeling in UWBGS device design.

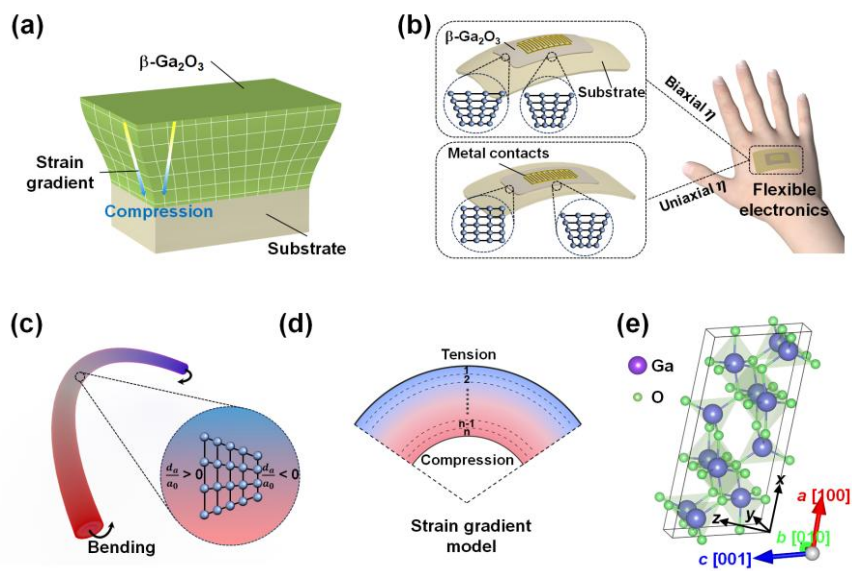

**Fig. 1. Non-uniform strain in nanostructures.** Schematics of (a) heteroepitaxial β-Ga₂O₃ film on substrate with biaxial strain ($\varepsilon$) and strain gradient ($\eta$) from lattice mismatch, (b) bent and warped flexible β-Ga₂O₃ photodetector exhibiting non-uniform strain with perpendicular $\eta$, (c) bent β-Ga₂O₃ nanowires under $\eta$. (d) Schematic of the strain gradient model, the elastic strain can be assumed to be homogeneous for each sub-cell $n$. (e) Conventional unit cell of β-Ga₂O₃ crystal structure. The $x$-$y$-$z$ coordinate system is defined to study the $\eta$ along the $z$-direction. $x$ and $y$ are the in-plane principal axes, and $z$ is the direction of the strain gradient.

## 2. Methods

## 2.1 First-principles-based neuroevolution machine-learning potential

The structural optimization of β-Ga₂O₃ is achieved using first-principles calculations within the framework of density functional theory (DFT) in the Vienna ab initio simulation package (VASP) [54]. The Perdew-Burke-Ernzerhof (PBE) exchange-correlation functional is employed, with the kinetic energy cutoff set to 520 eV and the total energy convergence criterion setting to $1\times 10^{-7}$ eV. To train the NEP model [52,55], ab initio molecular dynamics (AIMD) simulations are conducted to extract configurations from $2 \times 2 \times 2$ supercells over a temperature range of 100 K to 800 K. Uniform triaxial strains ranging from -3% to 3%, along with randomly generated small displacement structures, are introduced as structural perturbations. This strain range ensures the phase stability of β-Ga$_2$O$_3$ [41]. Uniaxial strain is simulated by stretching or compressing the lattice constant along the target (strained) axis while relaxing the lattice constants along the other two (unstrained) axes. Biaxial strain is modeled by simultaneously applying strain to the lattice constants along two axes, with the remaining axis relaxed. A 1 × 1 × 1 Γ-centered k-point grid is used to generate molecular dynamics trajectories, from which configurations are randomly extracted. The energies and interatomic forces are subsequently obtained via DFT static calculations, using an energy cutoff of 520 eV and 1 × 4 × 2 Γ-centered k-point grid [56].

After obtaining the training and testing database through first-principles calculations, active learning is then employed to identify unknown configurations via the farthest point sampling method, ultimately yielding 649 representative configurations—585 of which are used as the training set, while 64 are allocated as the testing set. As shown in Table I, during the training process, hyperparameters for radial ($r_c^R$) and angular $r_c^A$ cutoffs are set to 7 Å and 5 Å, respectively, with both radial ($n_{max}^R$) and angular ($n_{max}^A$) descriptor components set to 8. The maximum expansion orders ($l_{max}$) for three-body and four-body terms are set as 4 and 2, respectively. The number of neurons ($N_{neu}$) in the hidden layer is 50. Both weight $\lambda_1$ and $\lambda_2$ of the regularization terms are 0.05. The size of the population ($N_{pop}$) used by the separable natural evolution strategy is 50. A total generation ($N_{gen}$) of 250000 is applied.

**Table I.** The main hyperparameters used in the training of NEP.

| Hyperparameter | Value | Hyperparameter | Value |
|:---:|:---:|:---:|:---:|
| $r_c^R$ | 7 | $N_{neu}$ | 50 |
| $r_c^A$ | 5 | $\lambda_1$ | 0.05 |
| $n_{max}^R$ | 8 | $\lambda_2$ | 0.05 |
| $n_{max}^A$ | 8 | $N_{pop}$ | 50 |
| $l_{max}$ | 4, 2 | $N_{gen}$ | $2.5 \times 10^5$ |

## 2.2 Boltzmann transport equation calculations

In anharmonic lattice dynamics, *k* can be obtained by solving the linearized phonon BTE [50,57–59]:

$$k = \frac{1}{k_B T^2 \Omega N} \sum_\lambda n_\lambda^0 (n_\lambda^0 + 1)(\hbar \omega_\lambda)^2 \boldsymbol{v}_\lambda \otimes \boldsymbol{F}_\lambda, \qquad (1)$$

where $N_q$ is the number of discrete points in the Brillouin zone sampling grid. $\Omega$, $T$, and $k_B$ are the volume of a unit cell, absolute temperature in kelvin, and Boltzmann constant, respectively. $n_\lambda^0$, $\omega_\lambda$, and $\boldsymbol{v}_\lambda$ denote the occupation number, frequency, and group velocity of phonon mode $\lambda$, respectively. $\boldsymbol{F}_\lambda$ is the vector characterizing the deviation of phonon populations from equilibrium in the linear regime, expressed as:

$$\boldsymbol{F}_\lambda = \tau_\lambda (\boldsymbol{v}_\lambda + \Delta_\lambda), \qquad (2)$$

where $\Delta_\lambda$ is a linear combination of $\boldsymbol{F}_\lambda$ vectors. $\tau_\lambda$ is the phonon lifetime, whose inverse represents the phonon scattering rate. As demonstrated by Alkandari et al.'s work [46] and Tu et al.'s work [60], four-phonon interactions in β-Ga$_2$O$_3$ have little effect on thermal transport; therefore, only the lowest-order anharmonic three-phonon interactions are considered in this work.

The off-diagonal terms ($s \neq s'$) of the heat-flux operator ($k_{\text{off}}$), which describes the tunneling effect between different phonon branches $s$ and $s'$ can be calculated through [61,62]:

$$k_{\text{off}} = \frac{\hbar^2}{k_B T^2 VN} \sum_\lambda \sum_{s \neq s'} \frac{\omega_\lambda^s + \omega_\lambda^{s'}}{2} v_\lambda^{s,s'} v_\lambda^{s',s} \times \frac{\omega_\lambda^s n_\lambda^s (n_\lambda^s + 1) + \omega_\lambda^{s'} n_\lambda^{s'} (n_\lambda^{s'} + 1)}{\frac{4(\omega_\lambda^s - \omega_\lambda^{s'})^2}{\Gamma_\lambda^s + \Gamma_\lambda^{s'}} + (\Gamma_\lambda^s + \Gamma_\lambda^{s'})}, \qquad (3)$$

where $\Gamma_\lambda^s$ is the scattering rate of a phonon mode, and $n_\lambda^s$ represents the equilibrium Bose-Einstein distribution.

For bulk β-Ga$_2$O$_3$, BTE calculations incorporate $\tau_{3ph}^{-1}$ and $\tau_{iso}^{-1}$, while four-phonon scattering is neglected due to its weak contribution in β-Ga$_2$O$_3$ [46,60]. For thin films and nanowires, $\tau_b^{-1}$ is additionally included using Fuchs-Sondheimer (F-S) suppression functions [63–65] (see methods 2.5) rather than the oversimplified gray model $W/v^\alpha$, where $v$ is the group velocity, yielding different impact on in-plane and cross-plane *k* for thin films The Brillouin zone is sampled using a 5 × 17 × 9 mesh [47,60]. Since the normal process scattering rates are comparable than the Umklapp process scattering rates [32], the BTE is solved iteratively. To account for long-range interactions, the dielectric tensors and Born effective charges are coupled to the dynamic matrix [66]. All $k$ values include wave-like phonon tunneling contributions calculated from off-diagonal Wigner heat flux terms [61,62]. The main text focuses on the $k$ along the crystallographic axes ($k_a$, $k_b$, $k_c$), derived from the tensor's four independent components ($k_{xx}$, $k_{yy}$, $k_{zz}$, and $k_{xz}$, see SI Note II). After constructing the potential through machine learning method, the interatomic interactions required for BTE solution can be obtained quickly with a level of accuracy close to that of the first-principles method. Fig. 2 shows the entire computational framework.

## 2.3 Heat-averaged heat capacity, velocity, and lifetime

The heat-averaged [44] constant-volume heat capacity ($C_V$), $v$, and intrinsic phonon lifetime ($\tau_i$, excluding boundary scattering) are calculated by the following equations [46]:

$$\langle C_V \rangle = \frac{\sum_\lambda k_{\alpha,\lambda} |C_{V_\lambda}|}{\sum_\lambda k_{\alpha,\lambda}}, \tag{4}$$

$$\langle v \rangle = \frac{\sum_\lambda k_{\alpha,\lambda} |v_\lambda|}{\sum_\lambda k_{\alpha,\lambda}}, \tag{5}$$

$$\langle \tau \rangle = \frac{\sum_\lambda k_{\alpha,\lambda} |\tau_\lambda|}{\sum_\lambda k_{\alpha,\lambda}}, \tag{6}$$

where $\alpha$ represents the crystal orientation, and $\lambda$ is the mode index. The relative change in heat-averaged parameters (i.e. $\Delta\langle C_V \rangle/C_V, \Delta\langle v \rangle/v, \Delta\langle \tau \rangle/\tau$) and the relative change in $k$ ($\Delta k/k$) are calculated as the difference between the values of $\langle C_V \rangle$, $\langle v \rangle$, $\langle \tau \rangle$ and $k$ under strain and their corresponding values in the strain-free state, divided by the strain-free values.

## 2.4 Strain gradient model

$\eta$ induces lattice asymmetry, making strain gradient-induced scattering an intrinsic scattering mechanism. According to Matthiessen's rule, the total phonon scattering rate ($\tau_{tot}^{-1}$) is the sum of three-phonon scattering rate ($\tau_{3ph}^{-1}$), isotopic scattering rate ($\tau_{iso}^{-1}$), boundary scattering rate ($\tau_b^{-1}$), and strain gradient-induced scattering rate ($\tau_{sg}^{-1}$):

$$\tau_{tot}^{-1} = \tau_{3ph}^{-1} + \tau_{iso}^{-1} + \tau_b^{-1} + \tau_{sg}^{-1}. \tag{7}$$

The intrinsic phonon scattering ($\tau_i^{-1}$) includes $\tau_{3ph}^{-1}$, $\tau_{iso}^{-1}$, and $\tau_{sg}^{-1}$. As part of intrinsic phonon scattering, the $\tau_{sg}^{-1}$ can be quantized and expressed as [43]:

$$\tau_{sg}^{-1} = \alpha \times f \times \Delta\Theta, \tag{8}$$

where $f$ is the phonon Bose-Einstein distribution at a given temperature, $\Delta\Theta$ represents the phonon frequency shift induced by the strain gradient, and $\alpha$ denotes the ratio of the lattice vibration distance to the characteristic size ($W$) of the inhomogeneous strain field. The term $\alpha \times \Delta\Theta$ characterizes the phonon frequency perturbation. The vibration distance can be evaluated by calculating the atomic interaction cut-off distance $r_c$. Thus, $\alpha$ can be expressed as:

$$\alpha = \frac{r_c}{W}. \tag{9}$$

According to Fermi's golden rules, $\Delta\Theta$ corresponds to the transition rate between phonon mode $qj$ and $q'j'$ and can be computed as [67–71]:

$$\Delta\Theta = \sum_{q'j'} \frac{\pi\Omega}{\omega v_\lambda} |\langle q'j'|H(\omega)|qj\rangle|^2 \delta(\omega_{q'j'} - \omega_{qj}), \tag{10}$$

where $|qj\rangle$ represents the phonon eigenstate, $q$ and $j$ correspond to the wave vector and phonon polarization, respectively. This equation ensures energy conservation in the phonon scattering process. Under the weak perturbation of a strain gradient, the transfer matrix $H$ can be simplified using the Born approximation [69,70]:

$$H \approx H_{sg} = \frac{\phi_{sg} - \phi_0}{\sqrt{M_i M_j}}, \tag{11}$$

where $H_{sg}$ is the perturbation Hamiltonian, $M_i$ and $M_j$ denote the atomic masses of Ga and O, respectively. $\Phi_{sg}$ and $\Phi_0$ represent the harmonic force constants under inhomogeneous strain and strain-free states, respectively. The schematic of strain gradient model is shown in Fig. 1d.

To quantify $r_c$ in Eq. (8), the differential charge density [72] is computed with the atom displaced by 0.05 Å along the gradient direction. The resulting variation in charge density reflects the phonon interaction radius. In β-Ga$_2$O$_3$, the polar Ga-O ionic bonds and their dual coordination create long-range Coulomb interactions that lead to LO-TO splitting. Accordingly, $r_c$ is calculated to be 13.28, 15.77, and 12.34 Å along the directions orthogonal to the *ab, bc,* and *ac* planes, respectively (Fig. S2 in SI Note III).

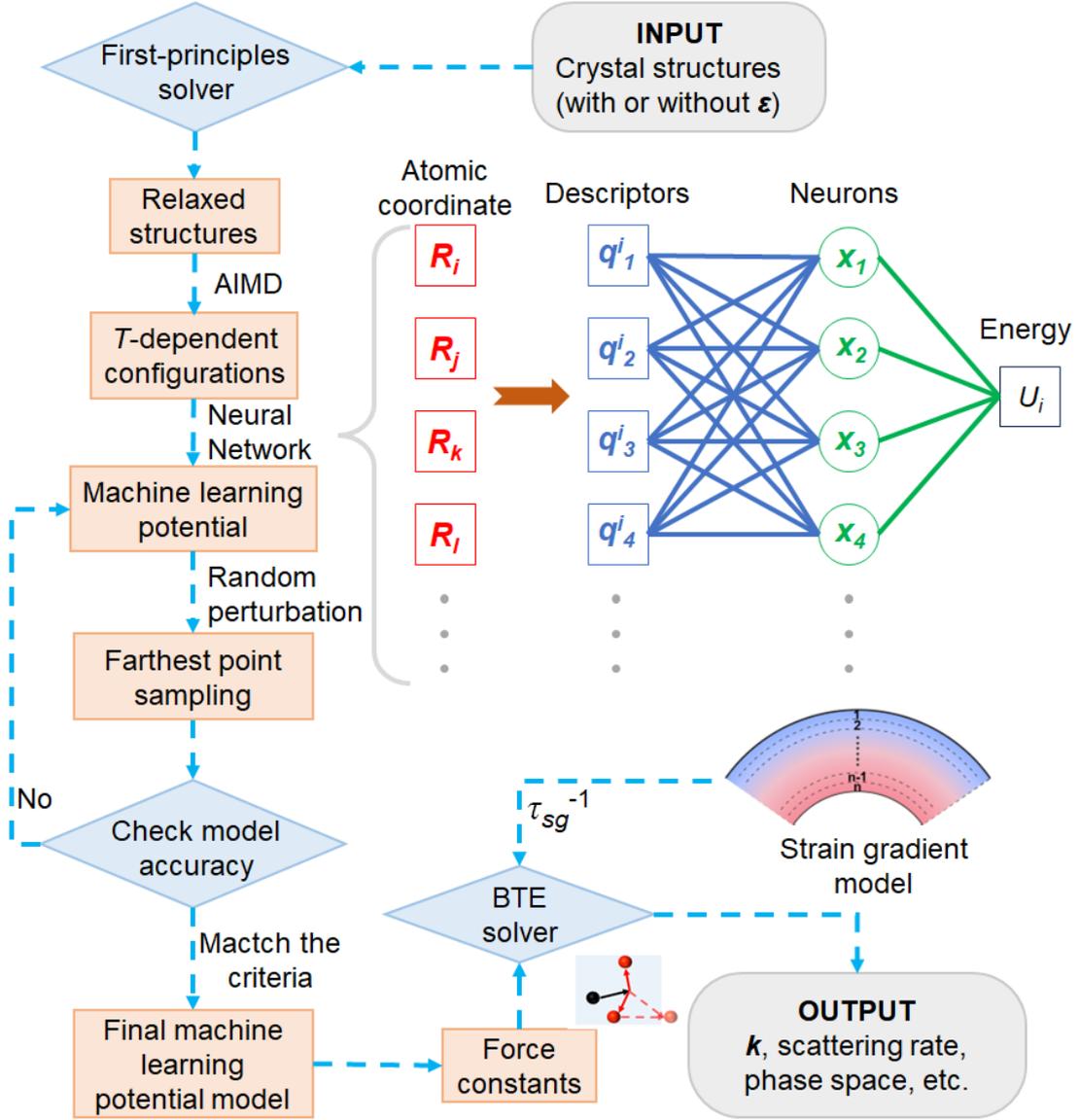

**Fig. 2.** The entire computational framework. The computational process extensively samples first-principles data, employing a separable natural evolution strategy and feedforward neural network [52] to construct the potential model, thereby significantly reducing the computational workload for calculating force constants under varying strains. Combined with strain gradient model, the strain gradient effect on thermal transport can be obtained.

## 2.5 Boundary scattering in nanofilms and nanowires

The BTE is solved using a modified version of the ShengBTE package [50], which accounts for boundary effects based on the suppression function [63,65,73–76]. The phonon lifetime in films ($\tau_{film}$) is expressed as:

$$\tau_{film} = F(\delta, p)\tau_{b,bulk} = F\left(\frac{d_s}{\Lambda_{bulk}}, p\right)\tau_{bulk}, \qquad (12)$$

where $\tau_{bulk}$ is the phonon lifetime in bulk materials, $F(\delta, p)$ is the reduction function, $\delta = d_s/\Lambda_{bulk}$, $d_s$ is the thickness of the film, $\Lambda_{bulk}$ is the mean free path within the bulk materials, $p$ is the fraction of phonons specularly reflected by the boundaries [73,75–77]. The $F(\delta, p)$ is calculated by:

$$F(\delta, p) = 1 - \frac{3(1-p)}{2\delta}\int_1^\infty \left(\frac{1}{\xi^3} - \frac{1}{\xi^5}\right)\frac{1-\exp(-\delta\xi)}{1-p\exp(-\delta\xi)}d\xi. \qquad (13)$$

The cross-plane $k$ ($k_\alpha$) in thin film is calculated through:

$$k_\alpha = \int_0^\infty (1 - \frac{\Lambda_\alpha}{d_s}\left(1 - e^{-\frac{d_s}{\Lambda_\alpha}}\right)) \times f(\Lambda_\alpha)d\Lambda_\alpha, \qquad (14)$$

where $\Lambda_\alpha = v_\alpha \tau$ is the mean free path along the transport $\alpha$ direction [75,76]. $f(\Lambda_\alpha) = C_V v_\alpha \Lambda_\alpha$ is the differential $k$ of bulk β-Ga$_2$O$_3$, where $C_V$ is the heat capacity.

Similar to the thin films, for purely diffuse scattering in a nanowire with an arbitrary cross section $A$, the reduction function is expressed as:

$$F(\Lambda_{bulk}, p=0) = 1 - \frac{3}{4\pi A}\int_A \int_0^{2\pi}\int_0^\pi \sin\theta \cos^2\theta \exp\left(\frac{-L_{OP}}{\Lambda_{bulk}}\right)d\theta d\phi dA, \qquad (15)$$

where $L_{OP}$ is the distance from a point $O$ on the cross section of the nanowire to a point $P$ on the surface of the nanowire along the direction given by the azimuthal angle $\theta$ and radial angle $\phi$ [73,77].

## 3. Results and discussion

### 3.1 Model construction and machine learning potential evaluation

We model strained β-Ga$_2$O$_3$ by applying incremental strains to its crystallographic axes and fully relaxing the internal atomic positions using constrained DFT as detailed in the methods section. Uniaxial strain ($\varepsilon^\alpha$) applied along a single axis $\alpha$ ($\alpha=a, b, c$) is $\varepsilon^\alpha = (x^\alpha - x_0^\alpha)/x_0^\alpha$, where $x^\alpha$ and $x_0^\alpha$ are the strained and unstrained lattice parameters, respectively. Biaxial strains ($\varepsilon^{\alpha\beta}$) are similarly created by applying strain concurrently along two axes. We define the effective $\eta$ along a Cartesian axis $n$ ($n=x, y,$ or $z$, as labeled in Fig. 1e) as $\eta_n^\alpha = \Delta\varepsilon^\alpha/W$, where $\Delta\varepsilon^\alpha$ is the difference in the strain component $\varepsilon^\alpha$ between opposing free boundaries separated by a distance $W$ (the film thickness or nanowire diameter). For all cases in this study, $\varepsilon$ is varied from -3% to 3% based on typical experimental feasibility [27–31,78,79].

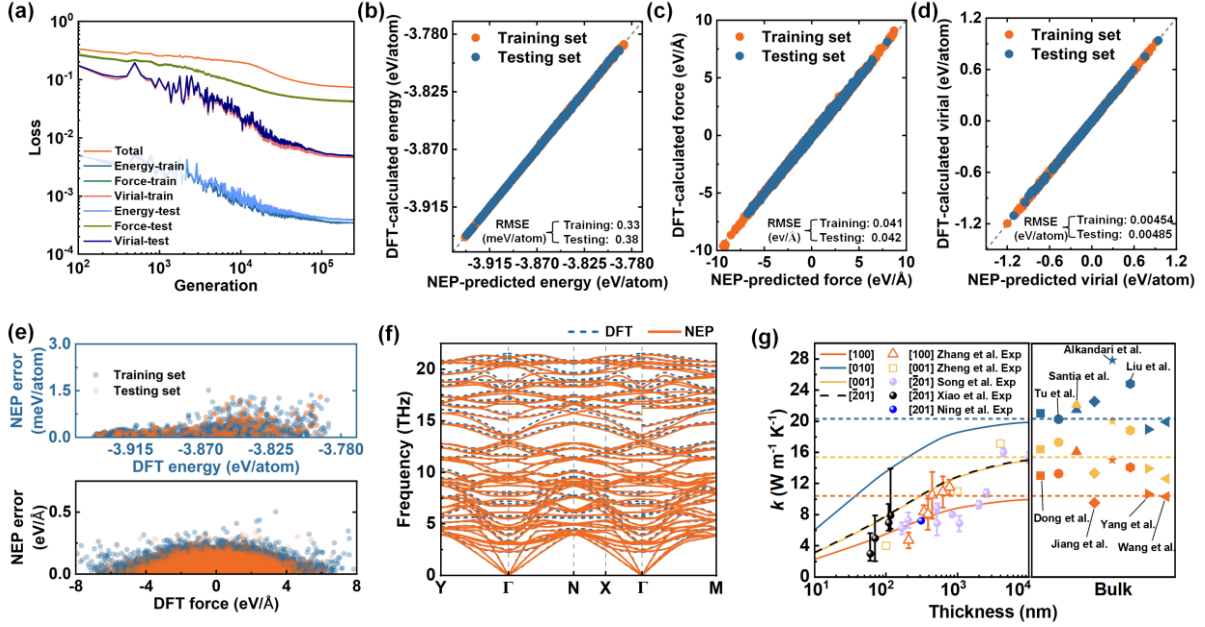

**Fig. 3. Training and evaluation of the neuroevolution potential.** (a) Evolution of the energy, force, and virial loss functions on the training and test sets during NEP training convergence. Comparison of (b) energy, (c) force, and (d) virial between NEP model predictions and DFT reference values. (e) Comparison of energy and force between NEP model predictions and DFT reference values. The error specifically represents the absolute error. (f) Phonon dispersion of strain-free β-$Ga_2O_3$ along the high symmetry path (Y-Γ-N-X-Γ-M) comparing DFT (dashed) and NEP (solid) calculations, demonstrating excellent agreement. (g) Components of $k$ tensor along the [100] ($a$-axis), [010] ($b$-axis), [001] ($c$-axis), and [-201] directions of β-$Ga_2O_3$ films (as illustrated in the crystal structure in panel (d)), shown as a function of thickness (left), and corresponding bulk values (right). The results from present calculations agree with previous experimental data [26,53,82–85] and theoretical results [46,47,60,89–91].

NEP employs a separable natural evolution strategy together with a feedforward neural network to construct the potential energy model (Fig. 2) [52], substantially reducing the enormous workload otherwise required to compute force constants at multiple strains due to the need for discretized strain gradients. The NEP is trained on these DFT-derived atomic configurations, with harmonic and anharmonic interatomic force constants extracted using the temperature-dependent effective potential [80,81] method for BTE calculations (details in methods). NEP training converges within 250,000 generations (Fig. 3a) with excellent reproduction of DFT energies, forces, virial (Fig. 3b-Fig. 3d), and phonon dispersions (Fig. 3f). The root mean square errors (RMSEs) of the energy (force) of the training and testing sets are 0.33 and 0.38 meV/atom (0.041 and 0.042 eV/Å), respectively. The root mean square errors (RMSEs) of the virial of the training (testing) set are 4.54 (4.85) meV/atom. Fig. 3e and Fig. S1 display the distribution of absolute errors in energy, force, and entropy results between NEP and DFT. The max relative error is less than 0.039% of the average energy. We also calculate $k$ along different crystal axes using the NEP, which show excellent agreement with experimental data [26,53,82–85] for both bulk ($k_a$: 10.4 W m$^{-1}$ K$^{-1}$, $k_b$: 20.3 W m$^{-1}$ K$^{-1}$, $k_c$: 15.4 W m$^{-1}$ K$^{-1}$ at 300 K) and thin-film β-Ga$_2$O$_3$ (Fig. 2g) [47,86,87]. The calculated lattice parameters ($a$=12.505 Å, $b$=3.10 Å, $c$=5.915 Å, and $β$=103.74°) are in agreement with the results from previous studies [32,41,46,47,60,86,88], demonstrating the robustness of our NEP model.

## 3.2 Thermal transport under uniform strain

To establish a baseline for subsequent analyses of $\eta$, we then examine the response of $k$ to uniaxial and biaxial $\varepsilon$ in bulk, 100-nm, and 10-nm thick films at 300 K (Fig. 4a-Fig. 4f). Our analysis reveals that the application of uniform strain produces substantial $k$ modulation in β-Ga$_2$O$_3$. This response is highly anisotropic, defined by two principal characteristics: each $k$ component shows maximum sensitivity to strain applied along its own crystallographic axis, and strains involving the $b$-axis induce the largest overall changes. This is exemplified under uniaxial strain, where $\varepsilon^b$ modulates its corresponding $k_b$ by -26.8%/+22.9% in 10-nm films, a significantly greater effect than the -18.6%/+13.3% modulation of $k_c$ by $\varepsilon^c$. The effect is amplified under biaxial strain, with $\varepsilon^{bc}$ of ±3% altering $k_c$ by as much as -48.3%/+22.4%. The full evolution of this thermal AR is detailed in Fig. 8. This strain sensitivity gradually increases in thicker films because $\tau_b^{-1}$ accounts for a smaller proportion of the $\tau_{tot}^{-1}$.

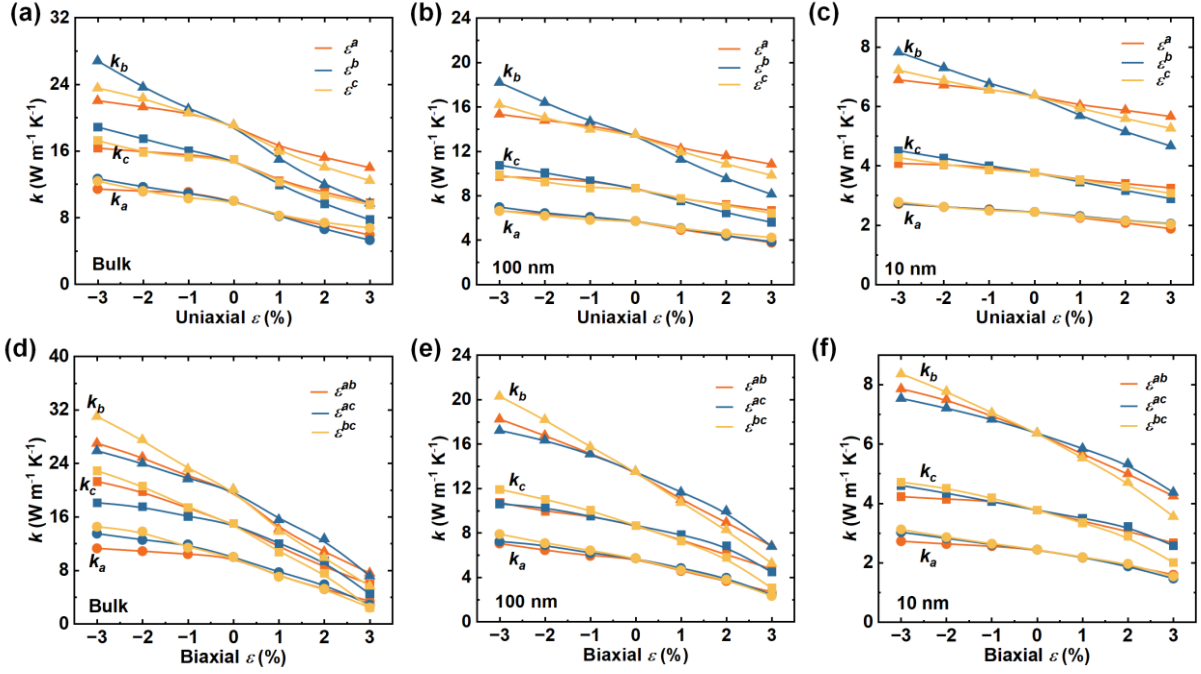

**Fig. 4. Thermal conductivity under uniaxial and biaxial ε.** $k$ of β-Ga$_2$O$_3$ under (a-c) uniaxial and (d-f) biaxial uniform strain at 300 K for (a,d) bulk, (b,e) 100-nm film, and (c,f) 10-nm film.

To identify the phononic origin of this high sensitivity, we analyze heat-averaged [44,46] constant-volume heat capacity ($C_V$), $v$, and intrinsic phonon lifetime ($\tau_i$, excluding boundary scattering) as functions of $k$ change ($\Delta k$) for all bulk β-Ga$_2$O$_3$ data from Fig. 4a-Fig. 4f. Across all strain scenarios, $C_V$ and $v$ remain largely invariant, whereas relative changes in $\tau_i$ scale nearly proportionally with $\Delta k$. As shown in Fig. 5a-Fig. 5c, in addition to the phonon frequency shift caused by ε, ε primarily affects the $C_V$ and $\tau_i$. Compressive ε increases $C_V$ and $\tau_i$, while stretching has the opposite effect. A quantitative analysis using Lin's concordance correlation coefficient [92,93] (CCC, labeled in Fig. 5d-Fig. 5f) confirms this exceptionally strong $\tau_i$-$k$ correlation. Specifically, the CCC values are 0.1484, 0.0465, and 0.8802 for $C_V$, $v$, and $\tau_i$, respectively, definitively showing that uniform strain modulates $k$ primarily through altering intrinsic three-phonon scattering rates rather than phonon stiffening/softening. The anisotropic and high sensitivity of $\tau_i$ to ε can be partly attribute to the strain-induced modifications of polar bonding within the β-Ga$_2$O$_3$ lattice [32,94].

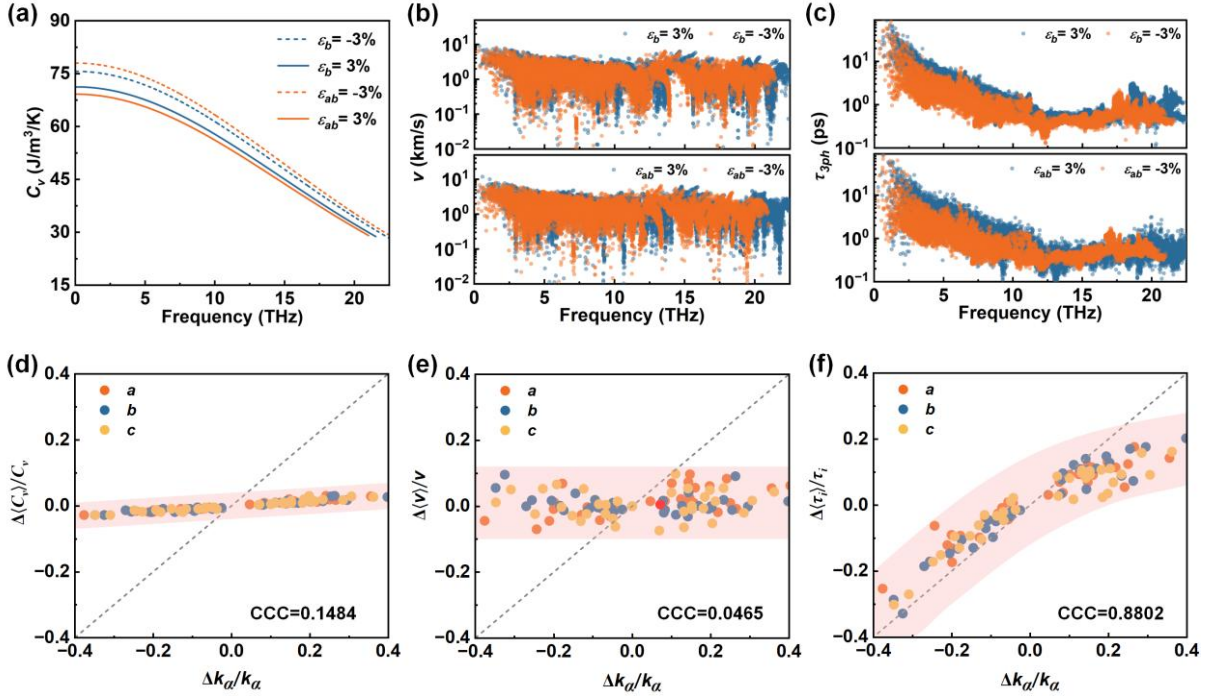

**Fig. 5. Relationship between relative changes in $C_V$, $v$, $\tau_i$ and $k$.** (a) Spectral $C_V$, (b) $v$, and (c) $\tau_i$ under uniaxial and biaxial $\varepsilon$ in bulk β-Ga$_2$O$_3$. Correlation between relative changes in $k$ ($\Delta k$) and heat-averaged (d) $C_V$, (e) $v$, and (f) $\tau_i$ for all crystallographic directions under uniaxial and biaxial $\varepsilon$ in bulk β-Ga$_2$O$_3$. Shadows in the image guide the viewer's gaze.

## 3.3 Thermal transport under strain gradient

To accurately study the $\eta$ effect on thermal transport, we first validate the strain gradient model (see methods) by calculating that the strain-induced broadening satisfies a linear relationship with the magnitude of strain. As shown in Fig. 6a, our calculated optical phonon (OP) mode shift at the Γ point (0.217 THz) under $\varepsilon$=1% agrees well experimental Raman measurements (0.195 THz) [40], confirming the accuracy of the NEP in capturing the phonon dispersion variation. Critically, we establish that phonon spectral broadening—quantified as the frequency difference between phonon modes at opposing boundary strains—scales linearly with the strain difference ($\Delta\varepsilon^\alpha$) between opposing boundaries across all crystallographic axes for both uniaxial and biaxial $\eta$ (Fig. 6b, Fig. S3 in SI Note IV). This linear scaling validates our approach based on the perturbation-quantized $\eta$ effect in Eq. (9). Furthermore, the pronounced broadening in OP (Fig. 6b) is particularly significant for evaluating the $\eta$ effect, as OPs contribute ~50% to $k$ in β-Ga$_2$O$_3$ [47]. These calculations also reveal the underlying anisotropic response of the material, where acoustic phonon frequency shifts under $\varepsilon^b$ and $\varepsilon^c$ exceed those under $\varepsilon^a$.

Fig. 6c shows that all $k$ components of β-Ga$_2$O$_3$ at 300 K are suppressed monotonically with increasing uniaxial $\eta$. A key finding emerges when comparing the impact of $\eta$ to that of a corresponding uniform strain. For a 100-nm film, the $k$ reduction from the relatively small $\eta_z^b$=0.06%/nm (corresponding to boundary $\varepsilon$=±3% for the 100-nm film) is comparable or smaller than that from the equivalent boundary tensile $\varepsilon^b$=3%. However, for a 10-nm film, this trend reverses: the reductions from $\eta_z^b$=0.6%/nm (31.8%, 35.3%, and 36.6% for $k_a$, $k_b$, and $k_c$, respectively) are significantly larger than those from the corresponding $\varepsilon^b$=3% (16.1%, 27.0%, and 23.4%). Substantial $\eta$-induced axial $k$ suppression can also be observed in nanowires architecture as shown in Fig. 8a and Fig. S4 (see SI Note V).

To understand the suppression of $k$ by $\eta$, we first investigate $\tau_{sg}^{-1}$ by mapping it onto the phonon dispersion at $\eta_z^b$=0.6%/nm for the 10-nm film (Fig. 6d). It reveals intense scattering when acoustic modes approach degeneracy with low-lying optical modes below the characteristic thermal frequency $f_p = \frac{k_B T}{2\pi\hbar} = 6.25$ THz at 300 K, leading to severe suppression of $k$. Mode-resolved analysis confirms that as $\eta$ increases, $\tau_{sg}^{-1}$ progressively dominates the total scattering rate for these crucial heat-carrying phonons below 10 THz (Fig. 6e). At $\eta_z^b$=0.6%/nm, the ratio $\tau_{sg}^{-1}/\tau_{tot}^{-1}$ exceeds 0.5 for most of these modes and approaches unity at the lowest frequencies, overtaking three-phonon scattering ($\tau_{3ph}^{-1}$) below 5 THz (Fig. S5a in SI Note VI). This is corroborated by the spectral $k_b$ in Fig. 6f, which shows the diminishing contribution of phonons below 10 THz as $\eta$ increases and indicates a selective suppression of long-wavelength phonons stemming partly from the disruption of long-range lattice order by $\eta$.

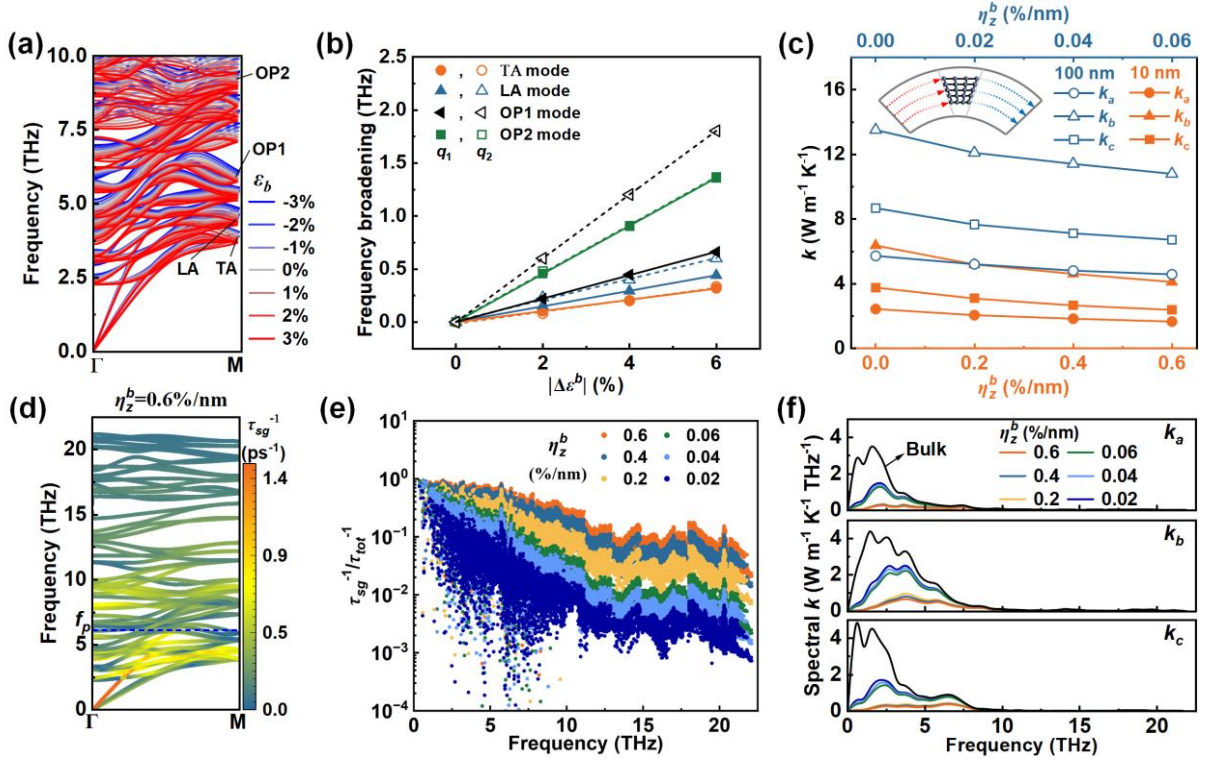

**Fig. 6. Thermal transport properties under $\eta_z^b$.** (a) Phonon dispersions of β-Ga$_2$O$_3$ under different $\varepsilon^b$. The selected TA (transverse acoustic), LA (longitudinal acoustic), and OP modes at $q_1$ point corresponding to the data in (b) are labeled. (b) The Phonon frequency difference broadening for TA, LA, and OP modes at selected $q_1$ (0.0, 0.5, 0.5) and $q_2$ (-0.5, 0.5, 0.0) under $|\Delta\varepsilon^b|$. The solid and dashed lines represent linear fitting results for $q_1$ and $q_2$, respectively. (c) Variation of $k$ in two kinds of β-Ga$_2$O$_3$ thin films (10 nm and 100 nm) under different values of $\eta_z^b$. The reductions in $k_a$, $k_b$, and $k_c$ are 20%, 20.1%, and 22.5% at $\eta_z^b$=0.06%/nm in the 100-nm film and 31.8%, 35.3%, and 36.6% in the 10-nm film under $\eta_z^b$=0.6%/nm, respectively. (d) Phonon dispersion mapped of $\tau_{sg}^{-1}$ under $\eta_z^b$=0.6%/nm. (e) Ratio of frequency-dependent $\tau_{sg}^{-1}$ to $\tau_{tot}^{-1}$ under different $\eta_z^b$. (f) Spectral $k_b$ at different $\eta_z^b$. All calculations are carried out at 300 K.

The impact of $\eta$ is even more pronounced under biaxial non-uniform strain ($\eta_z^{ab}$), a condition relevant to epitaxial films and warped flexible devices. At $\eta_z^{ab}$=0.6%/nm in a 10-nm film (corresponding to a boundary strain difference of $\Delta\varepsilon$=6% along both axes, Fig. 7a), the $k$ reductions (42.4%, 43.3%, and 41.6% for $k_a$, $k_b$, and $k_c$, respectively) are substantially larger than those from both uniaxial $\eta_z^b$ (Fig. 6) and biaxial $\varepsilon^{ab}$=3% (34.6%, 33.3%, and 29.1%). The variation of the $k$ tensor in different defined coordinate systems is shown in Fig. S6-Fig.

S8, exhibiting a similar decay pattern to $\eta_z^b$ and $\eta_z^{ab}$. This enhanced suppression stems from a broader activation of scattering processes across the phonon spectrum (Fig. S3). A comparison of scattering maps shows that while uniaxial $\eta_z^b$ primarily targets sub-thermal frequencies (Fig. 6d), biaxial $\eta_z^{ab}$ activates scattering across a wider range, strongly affecting both low-lying phonons and high-frequency optical modes between 7.5-10 THz (Fig. 7b and Fig. 7c). While the scattering rate ratio ($\tau_{sg}^{-1}/\tau_{tot}^{-1}$) also approaches unity at low frequencies, it crucially maintains a high value (>0.7) across the entire 0-10 THz range (Fig. 7d). Meanwhile, the $\tau_{sg}^{-1}/\tau_{3ph}^{-1}$ ratio (Fig. S5b) confirms the low-frequency dominance of $\tau_{sg}^{-1}$, demonstrating a more sustained and powerful scattering influence than the uniaxial case.

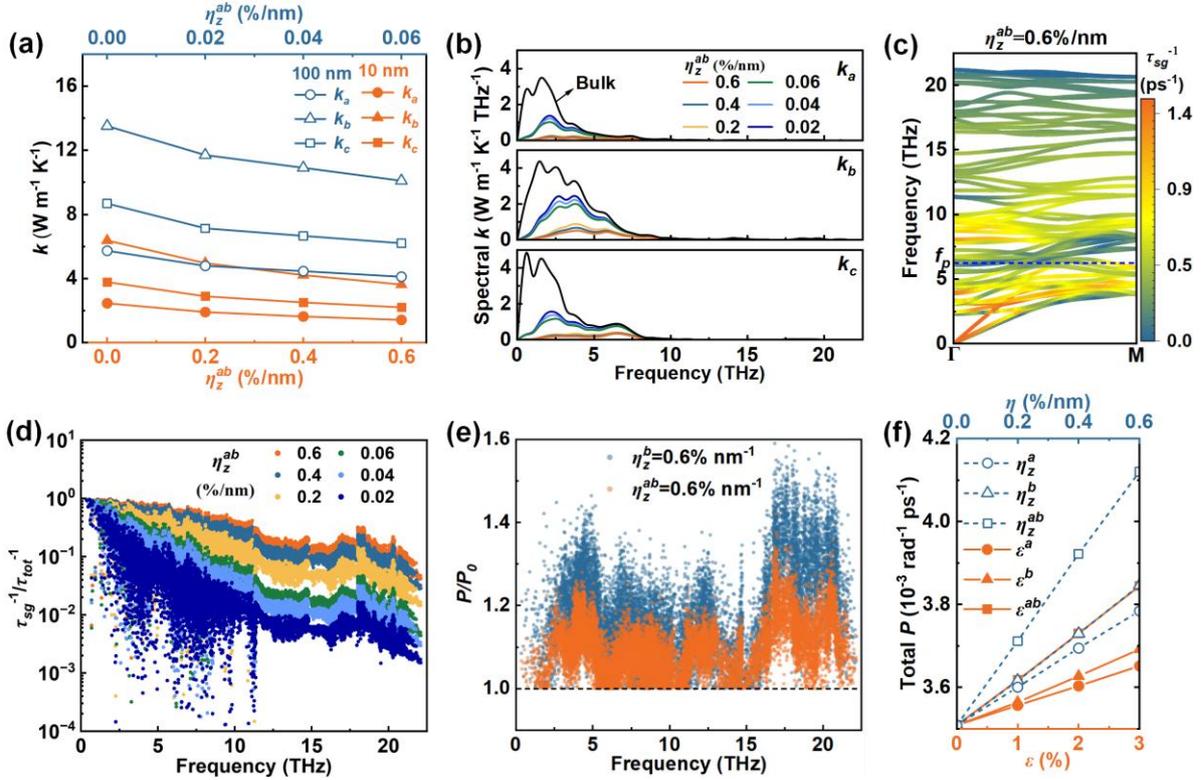

**Fig. 7. Thermal transport properties under $\eta_z^{ab}$.** (a) Variation of $k$ in β-Ga$_2$O$_3$ with two thicknesses (10 nm and 100 nm) under different values of $\eta_z^{ab}$. The reductions are 28.3% for $k_a$, 25.3% for $k_b$, 28.6% for $k_c$ at $\eta_z^b$=0.06%/nm in the 100-nm film, as compared to 42.4% for $k_a$, 43.3%, for $k_b$, and 41.6% for $k_c$ at $\eta_z^b$=0.6%/nm in the 10-nm film. (b) Spectral $k_b$ and $k_c$ under $\eta_z^{ab}$. (c) Phonon dispersion mapped by $\tau_{sg}^{-1}$ under $\eta_z^{ab}$=0.6%/nm. (d) Ratio of frequency-dependent $\tau_{sg}^{-1}$ to $\tau_{tot}^{-1}$ under $\eta_z^{ab}$. (e) Comparison of scattering phase space ($P$) at $\eta_z^b$=0.6%/nm and $\eta_z^{ab}$=0.6%/nm with the strain-free reference ($P_0$) in a 10-nm film. (f) Variation of total $P$ for different $\varepsilon$ and $\eta$ in the 10-nm film. All calculations are carried out at 300 K.

Fundamentally, the enhanced scattering under $\eta$ is driven by two interconnected mechanisms: (i) The breaking of translational symmetry, which broadens phonon spectra and relaxes momentum conservation rules, thereby activating forbidden scattering channels; (ii) The alteration of inter-mode frequency gaps due to the difference in energy shifts between phonon modes (Fig. 7c, Fig. 6b, Fig. S3), which modifies scattering dynamics. The combined effect alters the scattering selection rules, leading to a significant expansion of the phonon scattering phase space ($P$). Our calculations indicate that $P$ increases dramatically under both $\eta_z^b$ and $\eta_z^{ab}$ compared to the unstrained state ($P_0$) as shown in Fig. 7e. Critically, the growth rate of $P$ is faster under a $\eta$ than under the uniform $\varepsilon$ set equal to the gradient's boundary value (Fig. 7f). This effect is further amplified under biaxial $\eta_z^{ab}$, which produces the largest and fastest expansion of $P$. Notably, the slopes of $\eta$ and $\varepsilon$ along different axes also differ, with $\eta_z^b$ and $\varepsilon^b$ induced $P$ increase faster than other axes. This rapid expansion of available scattering channels provides the fundamental explanation for the potent and anisotropic suppression of $k$ by $\eta$.

$\beta$-$Ga_2O_3$'s exceptional response to $\eta$ stems from its fundamental structure: a monoclinic, polar lattice with strong LO-TO splitting and influential low-lying optical modes that are highly sensitive to symmetry breaking [32,94]. This allows its $k$ suppression to surpass that of leading material systems (Fig. 8a). Under a uniaxial $\eta_z^b$, the $k$ reduction in 10-nm films exceeds molecular dynamics predictions silicon nanocubes [44] and cubic BAs [45], while the reduction in 100-nm films is greater than that measured in silicon nanoribbons (34 nm × 85 nm) [43] and 3C-SiC [95]. In 10-nm and 100-nm nanowires, the reduction is slightly smaller, with the maximum decrease reaching 30.5% at $\eta_z^b$=0.6%/nm. The effect is magnified under a biaxial $\eta_z^{ab}$, which is unique to thin films, boosting suppression by a factor of ~1.5 compared to the corresponding $\varepsilon^{ab}$ and achieving a peak reduction of 43.3% for $k_b$ (Fig. 8d). Moreover, this response is anisotropic, as the sensitivities of $k_a$, $k_b$, $k_c$ and $k_{xz}$ differ with strain direction (see Fig. S6-Fig. S8 in SI Note VII), reflecting the complex interplay between polar bond strengthening, bond angle variations, and their effects on crystal symmetry and phonon anharmonicity [32].

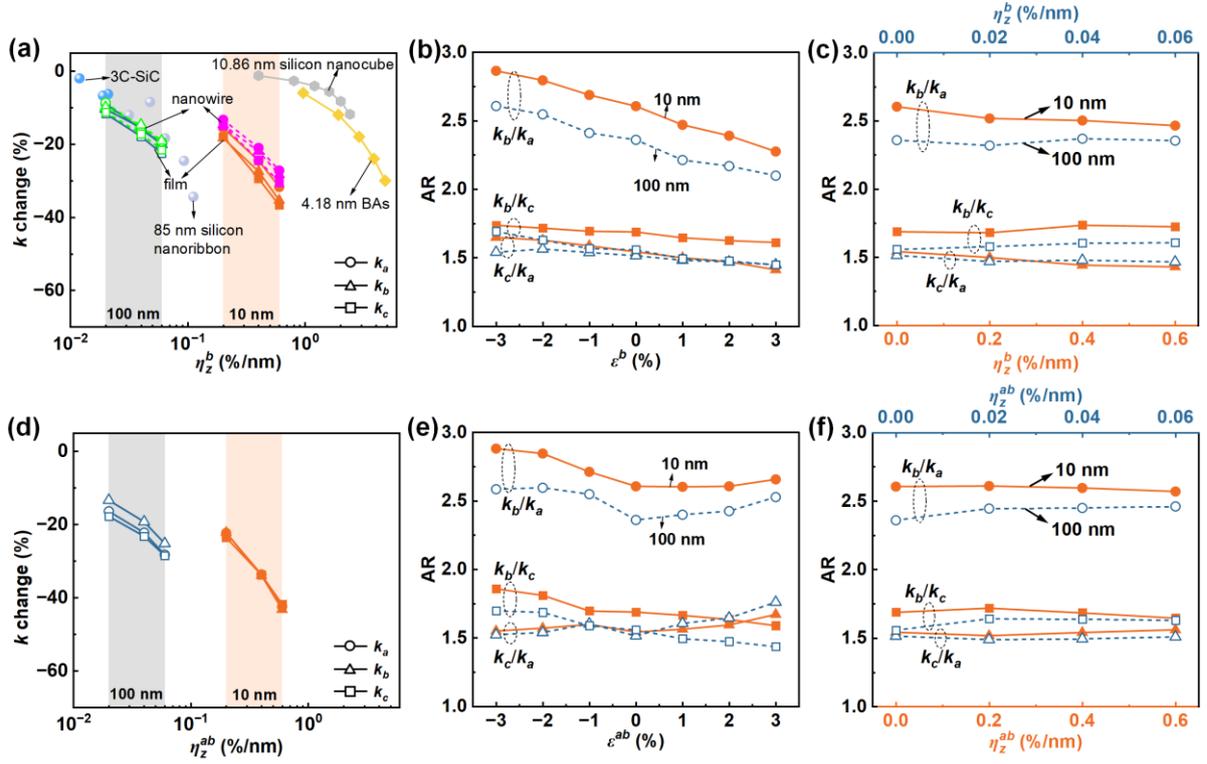

**Fig. 8. The comparison of *k* change and AR under $\varepsilon$ and $\eta$.** (a) Variation of *k* under uniaxial $\eta_z^b$ and comparison with the 3C-SiC, Si nanocube, Si nanoribbon and BAs results [43,45,95,96]. (b) Anisotropy ratios (ARs) for 10-nm (solid) and 100-nm (dashed) films under $\varepsilon^b$ (rectangle). (c) ARs for $k_b/k_a$ (circle), $k_b/k_c$ (square), and $k_c/k_a$ (triangle) under uniaxial $\eta_z^b$. With $\eta_z^b$ increasing from 0% to 0.6%/nm, the $k_b/k_a$ ratio in 10 nm (100 nm) films changes minimally from 2.60 (2.36) to just 2.47 (2.36). (d) Variation of *k* under $\eta_z^{ab}$. Under $\eta_z^{ab}$, the suppression effect is significantly enhanced, with a reduction rate ~1.5 times that under the equivalent uniform conditions. (e) ARs for 10-nm (solid) and 100-nm (dashed) films under $\varepsilon^{ab}$. (f) ARs for $k_b/k_a$ (circle), $k_b/k_c$ (square), and $k_c/k_a$ (triangle) under $\eta_z^{ab}$.

This anisotropic suppression of the individual *k* components provides a direct mechanism to engineer the thermal AR. However, The AR tuning result reveals a fundamental divergence between $\varepsilon$ and $\eta$ scenarios. Uniform $\varepsilon$ acts as a powerful lever for actively modulating the AR. As shown in Fig. 8b and Fig. S9 (see SI Note VIII), a uniaxial $\varepsilon^b$ applied to 10-nm films induces a near-linear tuning of the strongly reducing $k_b/k_a$, which transitions from a peak of 2.86 under -3% compression to 2.27 under +3% tension, due to the pronounced variation of $k_b$. Notably, the $k_b/k_c$ and $k_c/k_a$ ratios remain far more stable. This is due to the concurrent modulation of $k_c$ and $k_a$ by $\varepsilon^b$, which creates a compensating effect. The response to biaxial strain $\varepsilon^{ab}$ is more complex, revealing a counter-intuitive non-monotonic behavior (Fig. 8e): the $k_b/k_a$ ratio increases under both rising tensile and compressive strain, peaking at 2.88 under $\varepsilon^{ab}$=-3%, while $k_c/k_a$ and $k_b/k_c$ exhibit opposing trends. This demonstrates that uniform strain is a versatile tool for intentionally designing directional heat pathways. It's worth noting that the AR reduction at $\varepsilon^b$=2% is 13.2% for $k_b/k_a$, 4.7% for $k_b/k_c$, 8.7% for $k_c/k_a$, which is consistent with previously reported theoretical results (15.5% for $k_b/k_a$, 4.4% for $k_b/k_c$, and 9% for $k_c/k_a$) [32] in bulk β-Ga$_2$O$_3$, validating the accuracy of our first-principles-based machine learning potential approach.

In stark contrast, $\eta$ decouple the suppression of $k$ from the modulation of AR. As exemplified by both $\eta_z^b$ and $\eta_z^{ab}$ (Fig. 8c and Fig. 8f) and other conditions with gradients up to 0.6%/nm in Fig. S10, the AR remains remarkably stable even as $k$ is strongly reduced. With $\eta_z^b$ increasing from 0% to 0.6%/nm, the $k_b/k_a$ ratio in 10 nm (100 nm) films changes minimally from 2.60 (2.36) to just 2.47 (2.36). The stability of AR is preserved under biaxial $\eta_z^{ab}$. Compared with the unstrained case, a gradient of 0.6%/nm shifts the AR of $k_b/k_a$ in 10 nm (100 nm) films from 2.60 (2.36) to only 2.57 (2.46). This unique capability—suppressing heat flow without substantially altering intrinsic thermal anisotropy—distinguishes β-Ga$_2$O$_3$ from materials like nanoscale silicon [44]. From an application standpoint, this stability is advantageous, as it implies that process-induced strains like bending and warping will reduce overall $k$ without introducing unpredictable changes to a device's designed thermal anisotropy, ensuring robust performance modeling and optimization.

## Conclusion

In summary, by combining the first-principles-based machine learning potential with BTE and strain gradient model, this work reveals that strain gradients ($\eta$) in β-Ga$_2$O$_3$ produce fundamentally different and far more potent thermal suppression than uniform strain ($\varepsilon$). Under moderate uniaxial $\eta_z^b$=0.6%/nm, thin films and nanowires (10 nm) exhibit dramatic $k$ reductions of 32-37% and 27-30%, respectively, substantially exceeding $\varepsilon$ effects and surpassing leading benchmark materials including silicon and BAs. Biaxial gradients amplify this suppression, ~1.5-fold stronger than the corresponding uniform case, achieving peak reductions of 43.3% at $\eta_z^{ab}$=0.6%/nm—the strongest gradient-induced thermal suppression reported to date. Importantly, unlike $\varepsilon$, which varies thermal anisotropy ratios by ~25% across $\varepsilon^b$ from -3% to 3%, $\eta$ reduces $k$ without substantially altering thermal anisotropy, decoupling the two effects. This stability ensures process-induced strains predictably reduce heat flow without altering designed thermal pathways—a critical advantage for robust device optimization. The underlying physics reveals two synergistic mechanisms: gradient-induced symmetry breaking and enhanced mode coupling, which together activates previously forbidden phonon scattering, expanding the scattering phase space. These effects selectively target heat-carrying acoustic and low-lying optical phonons, with gradient-induced scattering rates dominating across 0-10 THz at moderate $\eta$. These findings establish $\eta$ as both a design challenge and an untapped opportunity in UWBGS thermal engineering, necessitating proper treatment in β-$Ga_2O_3$ device modeling while enabling targeted thermal isolation in flexible electronics, where non-uniform strains are ubiquitous.

## Conflicts of interest

The authors declare no competing financial interest.

## Author contributions

G.Z. and Q.Z. conceived the idea and supervised the project; G.Z. performed the programming and conducted the calculations; G.Z. and Q.Z. prepared the manuscript; G.Z., X.X., Z.Q., Y.X., and Q.Z. helped shape the research and analysis. S.Y., H.J., L.Y., Y.Z., and X.W. provided a critical review. Q.Z. and X.W. reviewed and revised the manuscript.

## Acknowledgements

We would like to acknowledge funding support from the National Key R&D Program of China (2024YFB4405700), the Nano & Material Technology Development Program through the National Research Foundation of Korea (NRF) funded by Ministry of Science and ICT (RS-2024-00444574), the Guangdong Basic and Applied Basic Research Foundation (2025A1515012898), the National Natural Science Foundation of China (No. 52576077), the Natural Science Foundation of Shandong Province (No. ZR2025MS874), the Youth Innovation Technology Support Program of Higher Education Institutions of Shandong Province (No. 2023KJ003), and the Qilu Youth Scholars Program of Shandong University. The authors thank the HPC Cloud Platform of Shandong University and HKUST Fok Ying Tung Research Institute and National Supercomputing Center in Guangzhou Nansha Sub-center for providing high performance computational resources. The authors would also like to thank the fund from Frontier Technology Research for Joint Institutes with Industry Scheme sponsored by the Center on Smart Sensors and Environmental Technologies at HKUST.

## Data availability

The data associated with the current study are available from the corresponding author on reasonable request.